\newcommand{\minpoint}{\mbox{$'\mskip-4.7mu.\mskip0.8mu$}}
\newcommand{\OII}{[O {\sc ii}]$\lambda3727$}
\newcommand{\OIII}{[O {\sc iii}]$\lambda$5007}
\documentstyle[emulateapj]{article}

\lefthead{Ajiki et al.}
\righthead{Lyman$\alpha$ Emitters at $z = 5.7$}

\begin{document}

\submitted{Accepted for publication in the Astronomical Journal}

\title{A Subaru Search for Lyman$\alpha$ Emitters at Redshift
5.7\altaffilmark{1, 2} }

\author{Masaru Ajiki\altaffilmark{3},
 Yoshiaki Taniguchi\altaffilmark{3},
 Shinobu S. Fujita\altaffilmark{3},
 Yasuhiro Shioya\altaffilmark{3},
 Tohru Nagao\altaffilmark{3},
 Takashi Murayama\altaffilmark{3},
 Sanae Yamada\altaffilmark{3},
 Kazuyoshi Umeda\altaffilmark{3},
 \&
 Yutaka Komiyama\altaffilmark{4}
 }

\altaffiltext{1}{Based on data collected at
 Subaru Telescope, which is operated by
 the National Astronomical Observatory of Japan.}
\altaffiltext{2}{Based on data collected at
        the W. M. Keck Observatory, which is operated as a scientific
        partnership among the California Institute of Technology, the
        University of California, and the National Aeronautics and Space
        Administration.}
\altaffiltext{3}{Astronomical Institute, Graduate School of Science,
 Tohoku University, Aramaki, Aoba, Sendai 980-8578, Japan}
\altaffiltext{4}{Subaru Telescope, National Astronomical Observatory of
Japan,
 650 North A'ohoku Place, Hilo, HI 96720}

\begin{abstract}

We present the results of a survey for Ly$\alpha$ emitters
at $z\approx 5.7$ based on optical narrow-band
($\lambda_{\rm c} = 8150$ \AA ~ and $\Delta\lambda = 120$ \AA),
and broad-band ($B$, $R_{\rm C}$, $I_{\rm C}$, and $z^\prime$)
observations of the field surrounding the high redshift quasar,
SDSSp J104433.04$-$012522.2, on the 8.2 m Subaru Telescope with
the Subaru Prime Focus Camera, Suprime-Cam.
This survey covers a sky area of
$\approx 720$ arcmin$^2$ and a co-moving volume of $\simeq
2 \times 10^5 ~ h_{0.7}^{-3}$ Mpc$^3$.
We have found 20 candidates of
Ly$\alpha$ emitters at $z \approx$ 5.7 with $\Delta z \approx 0.1$.
Two of them have been confirmed star-forming galaxies
at $z=5.655$ and $z=5.687$ from our follow-up optical spectroscopy.
We discuss star-formation properties of the 20 objects from a statistical
 point of view.
Our survey leads to a new estimate of the star formation rate density
at $z \approx 5.7$, $\sim 1.2 \times 10^{-3} h_{0.7} M_\odot$ yr$^{-1}$ Mpc$^{-3}$.

\end{abstract}

\keywords{cosmology: observations ---
   cosmology: early universe ---
   galaxies: formation ---
          galaxies: evolution}

\section{INTRODUCTION}

Lyman$\alpha$ emitters (LAEs) are one of the important populations of
star-forming galaxies in the early universe
(e.g., Partridge \& Peebles 1967; Larson 1974; Meier 1976).
Therefore, the search for LAEs at high redshift is essentially important for
the understanding of both the history of galaxies and the origin of
cosmic reionization (e.g., Loeb \& Barkana 2001).
Recent optical surveys with a narrowband filter have been
finding LAEs beyond redshift 5
thanks to the great observational capability of 8 -- 10 m class telescopes
(e.g., Hu, Cowie, \& McMahon 1998; Hu et al. 2002; Rhoads et al. 2003;
 Kodaira et al. 2003).

However, such surveys
have the following two technical limitations.
The first limitation comes from strong OH airglow emission lines
because they dominate at wavelengths longer than
7000 \AA ~ (e.g., Stockton 1999).
Although there can be seen several gaps at which there is little OH emission,
 this prevents us from finding very faint galaxies at high redshift.
The well-defined gaps appear around $\lambda \approx$ 7110 \AA, 8160 \AA,
and 9210 \AA. These gaps enable us to search for LAEs at
$z \approx$ 4.8 (Ouchi et al. 2003; see also Rhoads et al. 2000), 5.7
(Hu et al. 1998; Rhoads \& Malhtora 2001; Ajiki et al. 2002;
 Taniguchi et al. 2003), and 6.6 (Hu et al. 2002; Kodaira et al. 2003).
The second limitation is that survey volumes are so small because of
narrow filter band widths (e.g., $\sim 100$ \AA).

Although there are the above two limitations, recent successful results
encourage us to promote a new deep survey with a narrowband filter.
In order to gain survey volumes
and to reach faint limiting magnitude,
we need wide-field CCD cameras on 8 -- 10 m class telescopes.
The Suprime-Cam mounted at the prime focus of
the 8.2 m Subaru Telescope at Mauna Kea (Miyazaki et al. 2002)
 provides a unique opportunity for wide-field
 (a $34^\prime \times 27^\prime$ field of view),
 narrowband imaging surveys for emission-line objects at high redshift.
The efficiency of this instrument is higher by a factor of 30
than typical imagers mounted on the other 8 -- 10m class telescopes.\footnote{
Note that several wide-field cameras are also available on 4m class telescopes
 such as the Hale 5m telescope, KPNO and CTIO 4m telescopes, and CFHT
 (e.g. Roads et al. 2000; Stiavelli et al. 2001).
}
In this paper, we present results of our imaging survey using this
instrument to investigate observational properties of star-forming
objects at $z \approx 5.7$
(see also Ajiki et al. 2002; Shioya et al. 2002; Taniguchi et al. 2003).
Our survey field includes the quasar SDSSp J104433.04$-$012502.2
at $z=5.74$\footnote{
The discovery redshift was $z=5.8$ (Fan et al. 2000).
Since, however, the subsequent spectroscopic observations
suggested a bit lower redshift; $z=5.73$ (Djorgovski et al. 2001)
and $z=5.745$ (Goodrich et al. 2001), we adopt $z=5.74$ in this paper.}.

We adopt a flat universe with $\Omega_{\rm matter} = 0.3$,
$\Omega_{\Lambda} = 0.7$, and $h_{0.7}=1$ where $h_{0.7} = H_0/($70 km
s$^{-1}$ Mpc$^{-1}$). Throughout this paper,
magnitudes are given in the AB system.

\section{OBSERVATIONS AND DATA REDUCTION}

\subsection{Observations}

\begin{figure}
\epsscale{0.5}
\plotone{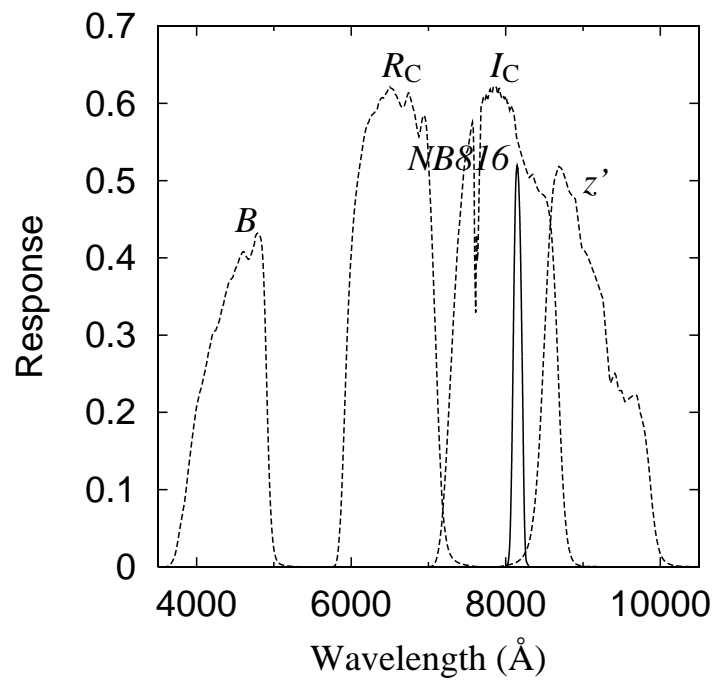}
\caption{Response curves (filter, optics, atmosphere transmission,
         and CCD sensitivity are taken into account)
         of the bands used in our observations.
\label{fil}}
\end{figure}

We have carried out a very deep optical imaging survey for
 faint LAEs in the field surrounding
 the quasar SDSSp J104433.04$-$012502.2 at redshift 5.74
 (Fan et al. 2000; Djorgovski et al. 2001; Goodrich et al. 2001),
 using the Suprime-Cam (Miyazaki et al. 2002)
 on the 8.2 m Subaru Telescope (Kaifu 1998) on Mauna Kea.
The Suprime-Cam consists of ten 2k $\times$ 4k CCD chips and provides
 a very wide field of view,
 $34^\prime \times 27^\prime$ (0.202 arcsec pixel$^{-1}$).
In this survey, we used a narrow-passband filter,
 {\it NB816}, centered on 8150 \AA ~ with a passband of
 $\Delta\lambda_{\rm FWHM} = 120$ \AA; the wavelength corresponds
 to a redshift of 5.65 -- 5.75 for Ly$\alpha$ emission\footnote{
         In Ajiki et al. (2002),
         the central wavelength of the {\it NB816} filter is given as
         $\lambda_{\rm center} =8160$ \AA. This should be read as
         $\lambda_{\rm center} =8150$ \AA.
         The redshift range covered by this filter is given as 5.663 -- 5.762.
         These values should be read as 5.65 -- 5.75.
         These changes do not affect any discussion in Ajiki et al (2002).
}.
We also used broad-passband filters,
 $B$, $R_{\rm C}$, $I_{\rm C}$, and $z^\prime$.
The total-response (filter, optics, atmosphere transmission,
 and CCD sensitivity are taken into account) curve of the each filter band
 used in our observations is shown in Figure \ref{fil}.

A summary of the imaging observations is given in Table \ref{tab:obs}.
All observations were done under photometric conditions,
 and the seeing was between 0$\farcs$7 and 1$\farcs$4
 during the observing run.
Photometric and spectrophotometric standard stars used in the flux
calibration
 are SA101 (Landolt 1992) for the $B$, $R_{\rm C}$, and $I_{\rm C}$ data,
 and GD 50, GD 108 (Oke 1990), and PG 1034+001 (Massey et al. 1988)
 for the {\it NB816} data.
The $z^\prime$ data were calibrated by using the $z^\prime$ magnitude of
 SDSSp J104433.04$-$012502.2 (Fan et al. 2000).
Since any quasar is a potentially variable object, the photometric
calibration
 of the $z^\prime$ data may be more unreliable than those of
 the other-band data.
Then, using colors of stars obtained from spectral energy distributions (SEDs)
of various spectral types of stars (Gunn \& Stryker 1983), we have
 made a small amount of correction ($+0.15$ mag) for the $z^\prime$ data.
The individual CCD data were reduced and combined using IRAF and
 the mosaic-CCD data reduction software developed by Yagi et al. (2002).

The total size of the reduced field is $32\minpoint08 \times 25\minpoint23
\approx 807$ arcmin$^2$.
After masking the regions contaminated by fringes and bright stars,
 our actual survey area is  $\approx$ 720 arcmin$^{2}$.
The final image of {\it NB816} is shown in Figure \ref{nbg} together with
actual survey area enclosed by solid line.
The volume probed by the {\it NB816} imaging has (co-moving) transverse
 dimensions of 4.0 $\times 10^{3} h_{0.7}^{-2}$ Mpc$^2$, and
 the FWHM of the filter correspond to a co-moving depth along
the line of sight of 44 $h_{0.7}^{-1}$ Mpc ($z_{\rm min} \approx 5.65$
 and $z_{\rm max} \approx 5.75$; note that the transmission curve of our
 {\it NB816} filter has a Gaussian-like shape).
Therefore, a total volume of $1.8 \times 10^{5} h_{0.7}^{-3}$ Mpc$^{3}$
 is probed in our {\it NB816} image.

\subsection{Source Detection and Photometry}

Source detection and photometry were performed using
 SExtractor version 2.2.2 (Bertin \& Arnouts 1996) double image mode.
A source is selected as a 13-pixel connection above the 2 $\sigma$ noise level
 on the {\it NB816} image.
Photometry is performed with a 2.8 arcsec diameter aperture for each band
 image of which image sizes are matched with the $R_{\rm C}$-band image (1.4
 arcsec).
The limiting magnitudes are ${\it NB816}=26.0$,
 $B=26.6$, $R_{\rm C}=26.2$, $I_{\rm C}=25.9$, and
 $z^\prime=25.3$ for a 3$\sigma$ detection with a 2.8 arcsec diameter
 aperture.
In the above source detection, we find $\sim 49,000$ sources down to
 ${\it NB816} = 26.0$.

\section{RESULTS}

\subsection{Selection of {\it NB816}-Excess Objects}

Since the central wavelength of the $I_{\rm C}$ filter is bluer than
 that of the {\it NB816} filter,
 we calculate a continuum magnitude at $\lambda = 8150$ \AA~
 that we refer to as the ``{\it Iz} continuum'', using a linear combination
 ($f_{\it Iz}= 0.76 f_{I_{\rm C}} + 0.24 f_{z^\prime}$)
 of the $I_{\rm C}$ and $z^\prime$ flux densities; a 3 $\sigma$ limit of
 ${\it Iz} \simeq 26.0$ in a 2.8 arcsec diameter aperture.
This enables us to sample more accurately the continuum at the same
 effective wavelength as that of the {\it NB816} filter.

We select {\it NB816}-excess objects using the following criteria;
\begin{eqnarray}
{\it NB816} & < & 25.0, \\
{\it Iz-NB816} & > & 0.9,
\end{eqnarray}
 where ${\it Iz-NB816} > 0.9$ correspond to $EW_{\rm obs} \gtrsim 180$ \AA.
In Figure \ref{nbcm}, we show the diagram between ${\it Iz-NB816}$ and
 {\it NB816} for the objects in the {\it NB816}-selected catalog.
There are 62 sources which satisfy the above two criteria.

\subsection{Selection of LAE Candidates}
\label{select}

Our main aim in the present survey is to find strong LAEs
at $z \approx 5.7$. However, strong emission-line sources at lower redshift
may be also found in our survey;
e.g., \OII ~ sources at $z \approx 1.2$,
H$\beta ~ \lambda$4861 sources at $z \approx 0.68$,
\OIII ~ sources at $z \approx 0.63$,
H$\alpha + $[N {\sc ii}] $\lambda$6583 sources at $z \approx 0.24$
 (see Fujita et al. 2003b),
 and so on.
In order to select LAEs at $z \approx 5.7$ from
 our emission-line objects, we apply the following criteria to all
emitters selected in the previous section;
\begin{eqnarray}
                 B  & > & 26.6,\\
R_{\rm C}-I_{\rm C} & > & 1.8~~~   {\rm for} ~~ I_{\rm C}  \leq  24.8, \\
          R_{\rm C} & > & 26.6 ~~  {\rm for} ~~ I_{\rm C}  >     24.8.
\end{eqnarray}
The first criterion means that objects at $z\approx 5.7$ must be undetected
at 3 $\sigma$ noise level in our $B$ band image.
Because the continuum break at
 $\lambda_0 = 912$ \AA ~(Lyman Break) shifts to
 $\lambda_{\rm obs} \approx 6200$ \AA, galaxies at $z \approx 5.7$
 are completely damped in $B$-band ($\lambda_{\rm max} \approx 5000$ \AA).
The second criterion is led as follows.
We show the diagram between $R_{\rm C}-I_{\rm C}$ and
 $I_{\rm C}-z^\prime$ for all the objects detected in the broad-band
 images in the right panel of Figure \ref{riz}.
We also show color evolutions of model galaxies with the SEDs typical to
SB (young starburst), Irr, Scd, Sbc, and E
as a function of redshift (left panel).
As for the generation of SEDs of model galaxies, we use GISSEL96 (Bruzual \& Charlot 1993).
The star formation rate (SFR) of model galaxies is proportional to
$\exp(-t/\tau)$, where we use $\tau=1$ Gyr.
Changing the age of model galaxies, we generate the SEDs of above five type
galaxies
($t=1$, 2, 3, 4, and 8 Gyr for SB, Irr, Scd, Sbc, and E, respectively).
We use the average optical depth derived by Madau et al. (1996)
 to estimate the absorption by intergalactic neutral hydrogen.
The model results show that LAEs at $z \sim 5.7$
may have $R_{\rm C}-I_{\rm C}\gtrsim 1.8$
and $I_{\rm C}-z^\prime \gtrsim 1.0$ although
the latter color constraint appears not so severe.
This figure also implies that lower-$z$ emitters could
 have $R_{\rm C}-I_{\rm C} < 1.8$.
However, the majority of our emitters are too faint
 ($I_{\rm C} > 24.8$) to be applied by this color criterion.
For such faint sources, we set another criterion; LAE candidates
 should be undetected above 2 $\sigma$ in the $R_{\rm C}$-band image
 (the third criterion).
By applying the above criteria to our 62 emitters, we select 20 LAE
 candidates at $z \sim 5.7$ and none of them has $I_{\rm C} \leq 24.8$.

Their positions and photometric properties are given in Table \ref{tab:LAE}.
It is noted that all of our LAE candidates are undetected above 3$\sigma$
level in the $B$- and $R_{\rm C}$-band images
 (i.e., $B>26.6$ and $R_{\rm C}>26.2$).
In Figure \ref{sum}, we show the $B$, $R_{\rm C}$, $I_{\rm C}$, {\it NB816},
 and $z^\prime$ images (upper panel)  and contour maps (middle panel)
 of the LAE candidates at $z \approx 5.7$.
The SED is also shown in the lower panel for each object.
We note that the two objects, No. 17 and No. 18, have been confirmed as galaxies
 at $z=5.655\pm0.002$ (Taniguchi et al. 2003) and $z=5.687\pm0.002$
 (Ajiki et al. 2002) by spectroscopic observations (see Section \ref{follow}).

\subsection{Contamination by Low-$z$ Emitters}

There is a possibility that
 some low-$z$ emitters are included in our final sample of the 20 LAE 
 candidates.
Since this affects our later discussion, we examine this possibility
 more quantitatively.

All the LAE candidates are very strong emission-line sources
 because they have $EW_{\rm obs} \gtrsim$ 180 \AA.
If they were lower-z emitters, 
 they would be star-forming galaxies at
 $z=0.24$, 0.63, or 1.19 corresponding to H$\alpha$,
 \OIII, or \OII, respectively.
They could be very strong emission-line galaxies because 
our observed equivalent width limit corresponds to
$EW($H$\alpha) \gtrsim$ 145 \AA,
$EW($[O {\sc iii}]) $\gtrsim$ 110 \AA, and
$EW($[O {\sc ii}]) $\gtrsim$ 82 \AA, respectively.
For the above three cases, we can nominally
 estimate the SFR for the 20 {\it NB816}-excess
 objects (i.e., the LAE candidates.). 
Then we can estimate their apparent $B$ and $R_{\rm C}$ magnitudes
based on galaxy models generated by using GISSEL96 (Bruzual \&
Charlot 1993). If these magnitudes are bright enough to be
detected in our $B$ and $R_{\rm C}$ images, we can reject the possibility
that they are low-$z$ emitters.

First, we estimate the nominal SFR for the above three cases; i.e.,
H$\alpha$, [O {\sc ii}], and [O {\sc iii}] emitters.
\begin{description}
\item{(1) }{The case of H$\alpha$ emitters:
We cannot estimate a pure H$\alpha$ line flux for each source from our data
because we have no information on the contribution of [N {\sc ii}]$\lambda 6583$
emission to the observed flux. Therefore, for simplicity, we adopt the average line flux ratio
for star-forming galaxies,  $f($H$\alpha)/f($[N {\sc ii}]$) = 2.3$,
obtained by Kennicutt (1992), and then estimate the pure H$\alpha$ flux
from our data.

We convert the H$\alpha$ luminosity to the SFR for each object,
 using the following relation;
\begin{equation}
\label{LHtoSFR}
SFR({\rm H}\alpha) = 7.9 \times 10^{-42} L({\rm H}\alpha)
 ~~~ M_\odot {\rm yr}^{-1}
\end{equation}
where, $L({\rm H}\alpha)$ is the H$\alpha$ luminosity in units
of ergs s$^{-1}$ (Kennicutt 1998).
In this estimate, the solar metallicity ($Z=Z_\odot=0.02$)
 and the Salpeter initial mass function with 0.1 -- 100 $M_\odot$ 
are assumed.
}
\item{(2) }{The case of [O {\sc ii}] emitters:
We use the following relation;
\begin{equation}
\label{LOIItoSFR}
SFR(\textrm{[O {\sc ii}]}) = 8.8 \times 10^{-42} L(\textrm{[O {\sc ii}]})
 ~~ M_\odot {\rm yr}^{-1}
\end{equation}
where $L$([O {\sc ii}]) is the [O {\sc ii}] luminosity
 in units of ergs s$^{-1}$.
In this estimate, we use equation (\ref{LHtoSFR}) together with
[O {\sc ii}] / H$\alpha = 0.9$ (Hippeline et al. 2003).
.
}
\item{(3) }{The case of [O {\sc iii}] emitters:
We use the following relation;
\begin{equation}
SFR(\textrm{[O {\sc iii}]}) = 1.0 \times 10^{-41} L(\textrm{[O {\sc iii}]})
 ~ M_\odot {\rm yr}^{-1}
\end{equation}
where $L$([O {\sc iii}]) is the [O {\sc iii}] luminosity 
in units of ergs s$^{-1}$.
In this estimate, we use equation (\ref{LOIItoSFR}) together with
 [O {\sc iii}]/[O {\sc ii}]$= 0.9$  (Hippeline et al. 2003).
}
\end{description}

\begin{figure}
\epsscale{0.61}
\plotone{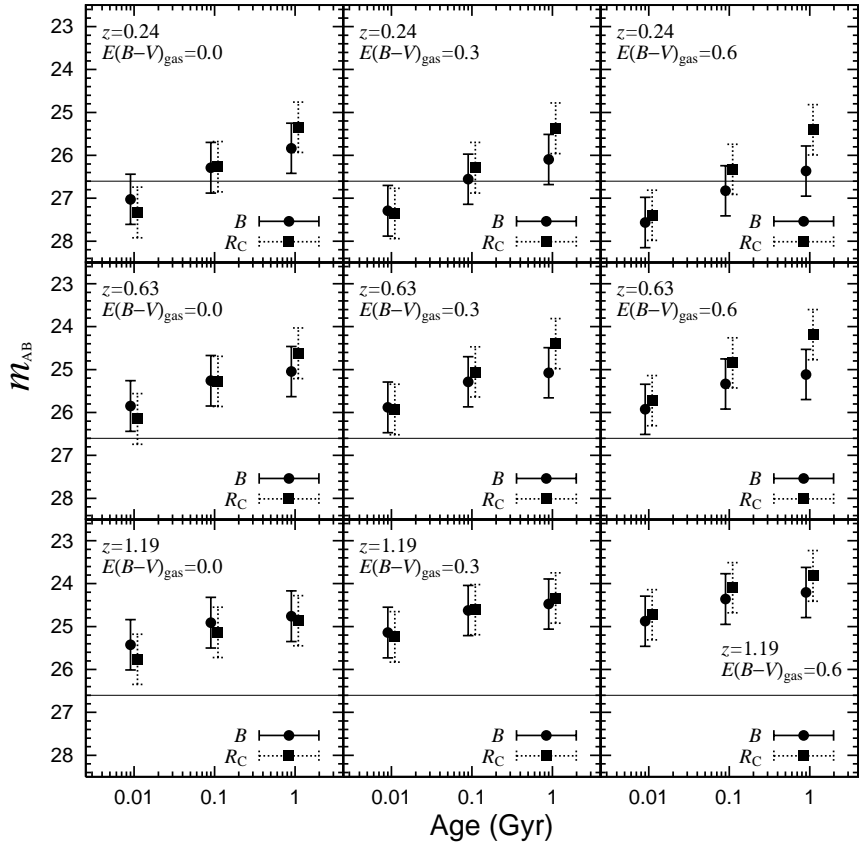}
\caption{$B$ and $R_{\rm C}$ magnitudes of the model galaxies at $z=0.24$,
         0.63, and 1.19 shown in three extinction cases
         ($E(B-V)_{\rm gas}=0$, 0.3, and 0.6).
         For readers' convenience, the age of each datapoint is shown
         slightly shifted from the real value (0.01, 0.1 and 1 Gyr);
         The $B$-band results are shifted to the left
         while the $R_{\rm C}$-band ones are to the right.
         The model galaxies are assumed to the star forming galaxies with
          a constant SFR estimated from the line fluxes of our LAE
          candidates.
         The vertical error bars correspond to
          the range of SFRs shown in Table \ref{tab:CONT}.
         The horizontal line shows our selection criteria of the LAE candidates
         (i.e., $B>26.6$ and $R_{\rm C}>26.6$).
\label{cont}}
\end{figure}

Next, we estimate the expected apparent $B$ and $R_{\rm C}$ magnitudes
 of model galaxies with the SFRs estimated above, given that objects
are located at the corresponding redshifts.
In order to estimate apparent $B$ and $R_{\rm C}$ magnitudes, we use GISSEL96
 assuming that these galaxies
 have been forming stars constantly for the recent past, 0.01, 0.1, or 1 Gyr
 and have solar metallicity ($Z=Z_\odot=0.02$).
The results are shown in the three left panels of Figure \ref{cont}
for the case of H$\alpha$, [O {\sc iii}], and [O {\sc ii}] emitters.

In the above estimates, we do not include the effect of internal extinction.
However, since the internal extinction affects both the line luminosity and the apparent
 magnitude, we examine how the extinction affects our results.
Since it is known that 
typical starburst galaxies have the color excesses for the nebular gas,
 $E(B-V)_{\rm gas}=0$ -- 0.6 (e.g., Calzetti et al. 2000; Cardiel et al. 2003),
we also estimate expected apparent $B$ and $R_{\rm C}$ magnitudes for
$E(B-V)_{\rm gas}=0.3$ and 0.6. 
In these estimates, we adopt
 the extinction law obtained by Calzetti et al. (2000).
The results for the cases of $E(B-V)_{\rm gas}=0.3$
 and $E(B-V)_{\rm gas}=0.6$ are shown in the second column and
 the right column of Figure \ref{cont}, respectively.

In Table \ref{tab:CONT}, we give a summary of our results.
As shown in Figure \ref{cont},  if our LAE candidates were low-$z$ emitters,
 they could be easily detected in our $B$ and $R_{\rm C}$ images.
Therefore, we can rule out the possibility that our LAE candidates
are low-$z$ emitters. This conclusion is hold even if the internal extinction
works  up to $E(B-V)_{\rm gas} = 0.6$.
The reason why the effect of the extinction does not play an important role is
that the obscuration by dust is more effective for the line emission from nebular
 gas than that for the stellar-continuum emission i.e.,
\begin{equation}
E(B-V)_{\rm star} = 0.44 E(B-V)_{\rm gas}
\end{equation}
 (Calzetti et al. 2000).
One remaining possibility seems that some candidates may be
H$\alpha$ emitters at $z \approx 0.24$ if they experience
a short-timescale star formation event
 (the duration must be shorter than $\sim 0.1$ Gyr).
However, it is unlikely that there are numerous such galaxies at $z=0.24$.
Therefore, we conclude that our LAE sample is free from the contamination 
 of low-$z$ emitters.

\subsection{Star Formers vs. Active Galactic Nuclei}

\begin{figure}
\epsscale{0.4}
\plotone{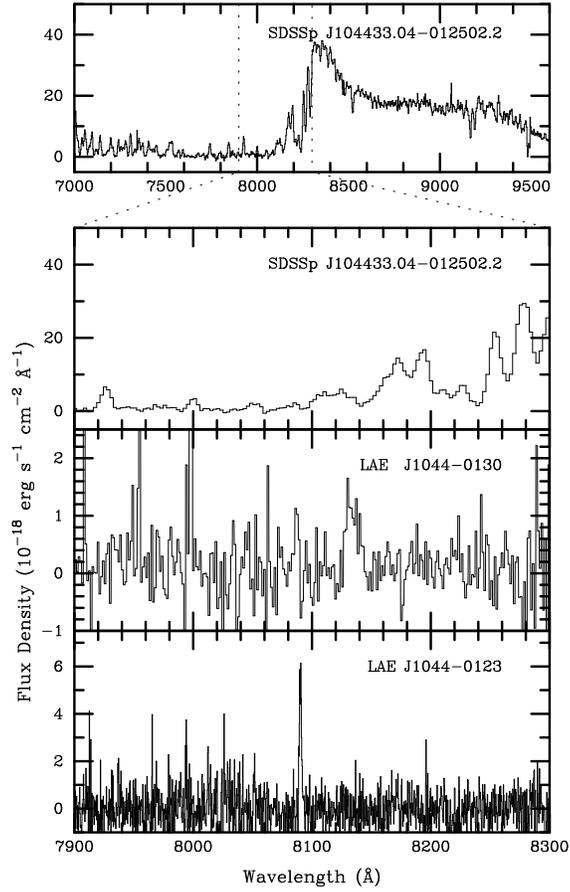}
\caption{
Optical spectra of SDSSp J104433.04$-$012502.2, LAE J1044$-$0130,
and LAE J1044$-$0123.
\label{sp3}}
\end{figure}

Previous narrowband imaging surveys for high-$z$ emission-line sources
 resulted in the discovery of star-forming galaxies (e.g., Hu et al. 1998,
 2002; Kodaira et al. 2003; Rhoads et al. 2003).
However, it is uncertain whether our LAE candidates are genuine star-forming
 galaxies or active galactic nuclei, especially quasars, although our two LAEs
 are star forming galaxies (Ajiki et al. 2002; Taniguchi et al. 2003).
Our survey is biased to detect very strong emission-line objects
($EW_{\rm obs} \gtrsim 180$ \AA). Therefore, if observed equivalent widths
of Ly$\alpha$+N {\sc v} emission lines are smaller than the EW limit,
quasars are not selected as LAE candidates. For example,
SDSSp J104433.04$-$012502.2 at $z=5.74$ (Fan et al. 2000) and 
SDSS J114816.64+525150.2 at $z=6.43$ (Fan et al. 2003) have 
$EW_{\rm obs} \sim 180$ \AA\footnote{
          Note that the peak wavelength
          of the Ly$\alpha$+N {\sc v} emission of SDSSp J104433.04$-$012502.2
          is $\simeq 8300$ \AA ~ (see Figure \ref{sp3}).
          Therefore only a part of the Ly$\alpha$ flux is probed in our
          {\it NB816} imaging, and thus this quasar is not selected as a LAE
          in our survey.}.
However, typical quasars have
 $EW_0 \sim 70$ \AA ~ (e.g., Fan et al. 2001).
In fact, some very high-$z$ quasars around $z \sim 6$ have 
 $EW_0 \sim 70$ \AA ~ (e.g., Fan et al. 2002, 2003).
If such quasars at $z \approx 5.7$ are present in our survey field,
 they could be selected as LAE candidates because $EW_{\rm obs}$ exceeds 450 \AA.
Although broader line widths of Ly$\alpha$+N {\sc v} emission lines make it
 difficult to detect high-$z$ quasars in any narrowband imaging surveys,
 it seems important to estimate a probable fraction of quasars in our LAE sample.

For this purpose, we estimate the expected number of quasars,
 $N_{\rm quasar}$, in our survey volume.
In the estimate, we assume that the number density of quasars at $z\approx5.7$
follows the maximum likelihood solution derived by Fan et al. (2001)
based on their data of 39 luminous quasars at $z=3.6$ -- 5.0;
\begin{eqnarray}
\log \Phi (z,<M_{1450})&=&(-7.31 \pm 0.19) \nonumber\\
                       & &-(0.47 \pm 0.15)(z-3) \nonumber\\
                       & &+(0.63 \pm 0.10)(M_{1450}+26),
\end{eqnarray}
where $\Phi (z, <M_{1450})$ is the number density of quasars brighter than
 $M_{1450}$ at redshift $z$ in units of Mpc$^{-3}$ and $M_{1450}$ is the absolute
 magnitude of the quasar continuum in the rest frame at 1450\AA.
Using this equation, we estimate the number density of quasars,
 $\log \Phi(5.7, <-22.0) \simeq -6.06 \pm 0.60$
 and the number of quasars in our survey volume,
 $N_{\rm quasar} \sim 0.16^{+0.49}_{-0.12}$,
 where $M_{1450}=-22.0$ corresponds to $z^\prime \approx 26.8$
 and $EW_0\approx 70$ \AA, if ${\it NB816}=25.0$.
In this estimate, we use a $\Lambda$ model with $\Omega_{\rm matter} = 0.35$,
 $\Omega_{\Lambda} = 0.65$, and $H_0=65$ km s$^{-1}$ Mpc$^{-1}$,
 following the manner of Fan et al (2001).
We note that the survey volume for quasars with the broad Ly$\alpha$
 + N {\sc v} emission lines is larger than the value using above estimate.
The survey volume for such quasars could increase by a factor of $\sim 3$.
We also note that the faint end of the optical luminosity function of
 the quasars is expected to be much flatter than the bright end
 which we use in the above estimate (see Boyle et al. 2000).
If this is the case, the expected number of quasars in our survey volume
 is further lowered.
In conclusion, the expected number of quasars in our LAE candidates
 is only one at most.

Note that Malhotra et al. (2003) find no X-ray sources
 associated with 49 LAEs at $z=4.5$ or 5.7.
It is also noted that the results of Steidel et al. (2002)
 show that only 29 of 988 LBGs at $z\sim$ 3 
 have AGN-like emission lines; i.e., $\simeq 3$ \%.
These results appear to be consistent with our estimate given above.

\subsection{Spatial Distribution of Emitter Candidates}

\begin{figure}
\epsscale{0.45}
\plotone{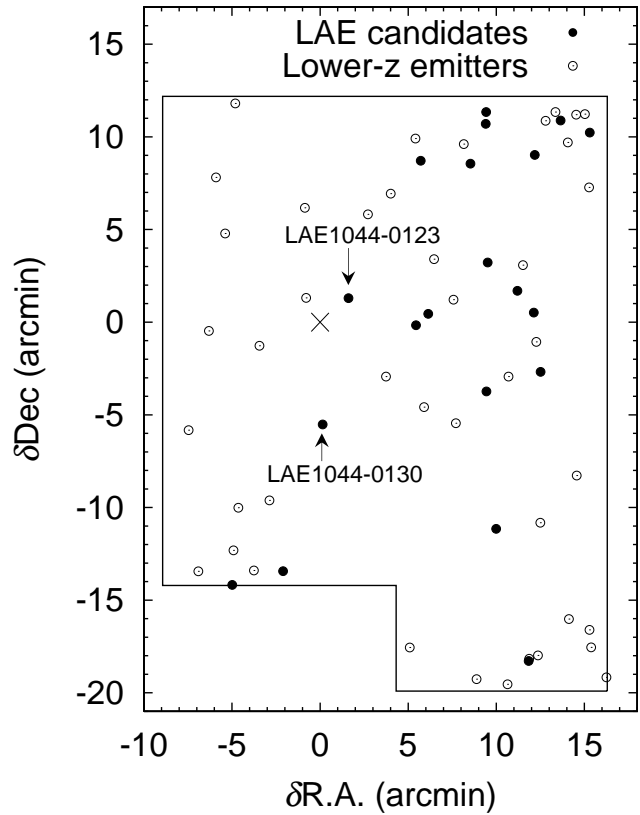}
\caption{Celestial positions of the 20 LAE candidates
         at $z \approx 5.7$, and 42 lower-$z$ emitter candidates.
         The boundary of survey area is also shown (solid line).
         The position of SDSSp J104433.04$-$012502.2 is shown by
         ``$\times$''.
         The two LAEs at $z\approx 5.7$ confirmed by our spectroscopy
         are marked.
\label{map}}
\end{figure}

We show the spatial distributions of the LAE candidates at $z \approx 5.7$.
In Figure \ref{map}, we plot the spatial distributions for our LAE
 candidates.
The 20 LAE candidates appear to be located at west of
 SDSSp J104433.04$-$012502.2, while the low-$z$ emitter candidates are
 distributed randomly in the sky.
However, we do not discuss further detail on this issue
 because spectroscopic confirmation has not yet been completed.

\section{DISCUSSION}

\subsection{Rest-Frame Ly$\alpha$ Equivalent Widths of the LAE candidates at
$z\approx 5.7$}

\begin{figure}
\epsscale{0.4}
\plotone{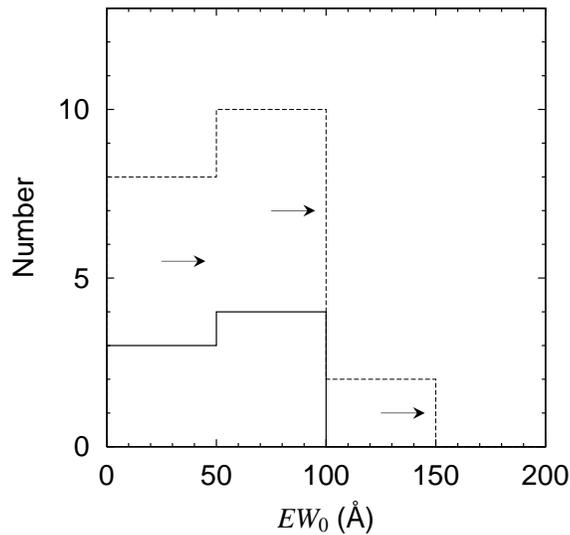}
\caption{Histogram of $EW_0$(Ly$\alpha$) for our 20 LAE candidates.
         The solid histogram is for seven LAE candidates
         with the continuum detected above $3\sigma$
         (i.e. {\it Iz}$\leq$ 26.0).
         The broken histogram is for all our LAE candidates,
         where $EW_0$ of LAE candidates with undetected continuum are
         estimated as lower limits assuming ${\it Iz}=26.0 ~(3\sigma)$.
\label{ewhist}}
\end{figure}

First, we investigate the rest-frame Ly$\alpha$ equivalent widths,
 $EW_0$(Ly$\alpha$), of our 20 LAE candidates (see Table \ref{tab:LLAE}).
In Figure \ref{ewhist}, we show a histogram of $EW_0$(Ly$\alpha$) for the 20
 sources.
Eleven of the 20 LAE candidates are undetected above 3$\sigma$ level in
 {\it Iz}.
For these LAE candidates,
 we estimate lower limits assuming ${\it Iz}=26.0 ~(3\sigma)$.
It is shown that the rest-frame equivalent widths range from
 25\AA ~ to $>120$ \AA.
This range is consistent with that of LAEs at $z>5$
 (e.g., Dey et al. 1998; Hu et al. 2002; Dawson et al. 2002).

\subsection{Space Density of the LAE Candidates at $z \approx 5.7$}

\begin{figure}
\epsscale{0.5}
\plotone{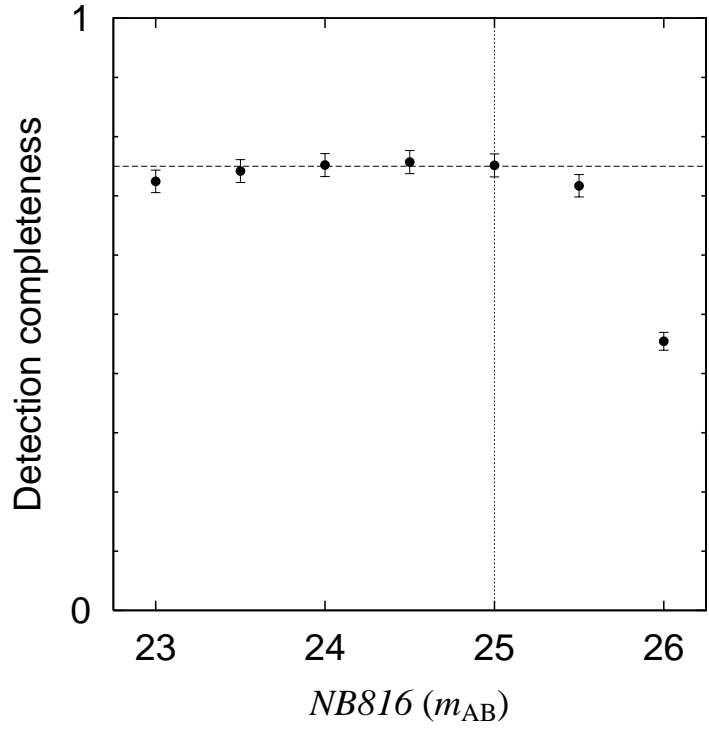}
\caption{Detection completeness of LAEs as a function of
         the apparent {\it NB816} magnitude derived from the simulation.
         The horizontal line shows the detection completeness of 0.75.
         The vertical line shows the detection limit of our survey
         (${\it NB816}=25.0$).
         \label{dc}}
\end{figure}

Since we have detected the 20 LAE candidates at
 $z \approx 5.7$ objects in the volume of $1.8 \times 10^{5} h_{0.7}^{-3}$
 Mpc$^{3}$, we obtain the space density of the LAE candidates at
 $z \approx 5.7$, $n({\rm Ly}\alpha) \simeq 1.1 \times 10^{-4} h_{0.7}^{3}$Mpc$^{-3}$.
Because the above estimate needs correction of the probability of detection,
 we have performed a simulation using the IRAF ARTDATA.
We assume that emitters have little continuum and are point sources.
We generated 2,000 model emitters for each magnitude interval
 ($\Delta m = 0.5$ mag).
After smoothing the model-emitter images
 to match to the image size of our final image data,
 these model emitters are put into the final image data.
Then we try to detect them using SExtractor and to select as a LAE with
 the same procedure as that used in our paper except for the magnitude
 criterion (${\it NB816}<25.0$).
The detection completeness of the model galaxies is shown in
Figure \ref{dc} as a function of AB magnitude.
From this result, the detection completeness is nearly constant at 0.75 for
 ${\it NB816} <25.0$ and then decreases for ${\it NB816} >25.0$.
This analysis shows that our detection is fairly well done for
 ${\it NB816} <25.0$ although we have to correct for the detection
 completeness factor when we discuss the star formation rate density.
Our analysis shows that the detection completeness is
 0.75 even for brighter sources.
The reason for this is that a large number of foreground objects affect
 the detection procedure in the broad band images (i.e., $B$ and $R_{\rm C}$)
 even though our {\it NB816} data are deep enough to
 detect emitters with ${\it NB816} <25.0$.
After correcting for the detection completeness, a space density of
 the LAE candidates is
 $n({\rm Ly}\alpha) \simeq 1.5 \times 10^{-4} h_{0.7}^{3}$ Mpc$^{-3}$ at $z \approx 5.7$.
On the other hand, Rhoads \& Malhotra (2001) obtained
$n({\rm Ly}\alpha) \simeq 8.2 \times 10^{-5} h_{0.7}^{3}$ Mpc$^{-3}$ at
$z \approx 5.7$
 for their full sample (18 LAEs; see also Rhoads et al. 2003)\footnote{
          $n({\rm Ly}\alpha) \simeq 5.9 \times 10^{-4} h_{0.7}^{3}$ Mpc$^{-3}$
          for their reduced sample (13 LAEs).}.
This density is lower by a factor of 2 than ours.
More accurately, we need compare the space density of the LAE candidate in 
the same luminosity range since our survey limit of Ly$\alpha$ luminosity
is higher than that of Rhoads \& Malhotra (2001).
When their LAE candidates fainter than
 the faintest Ly$\alpha$ luminosity in  our sample,
$5.0 \times 10^{42} h_{0.7}^{-2}$ ergs s$^{-1}$, are not taken into consideration,
the space density of their LAE candidates decreases to 
 $n({\rm Ly}\alpha) \simeq 4.5 \times 10^{-5} h_{0.7}^{3}$ Mpc$^{-3}$.
This density is lower by a factor of 3 than ours.
This difference may be partly due to that their value is not corrected
 for the detection completeness.
However, this correction seems too small to explain the difference.
Therefore, it is possible that the difference of the observed fields
 between the two studies (i.e., the cosmic variance). 

\subsection{Ly$\alpha$ Luminosity of the LAE Candidates}

\begin{figure}
\epsscale{0.5}
\plotone{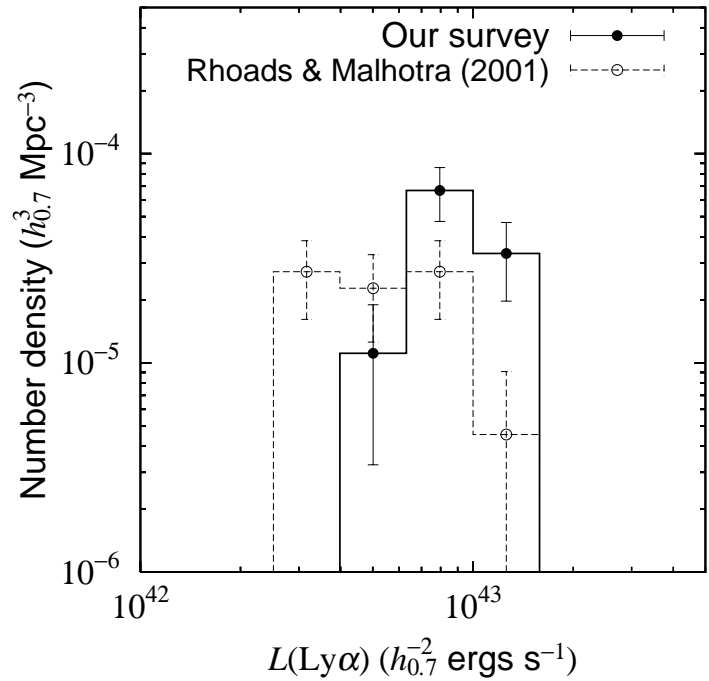}
\caption{Distribution of Ly$\alpha$
         luminosity is compared to that derived by
         Rhoads \& Malhotra (2001).
         We note that Ly$\alpha$ luminosities for their LAE candidates are
         estimated from the flux densities shown in Fig. 1 of
         Rhoads \& Malhotra (2001).
\label{ll}}
\end{figure}

We investigate the Ly$\alpha$ luminosities, $L$(Ly$\alpha$),
 of the $z \approx 5.7$ candidates.
We assume that all the sources have a redshift of 5.70
 which corresponds to the case that the Ly$\alpha$ is shifted to
 the central wavelength of the {\it NB816} filter.
In Table \ref{tab:LLAE}, we give the Ly$\alpha$ luminosities for
 the 20 LAE candidates.
The derived Ly$\alpha$ luminosities range from
$\approx 5 \times 10^{42} h_{0.7}^{-2}$ ergs s$^{-1}$
 to $\approx 1.4 \times 10^{43} h_{0.7}^{-2}$ ergs s$^{-1}$.
In Figure \ref{ll}, we show the distribution of Ly$\alpha$ luminosities for
 our 20 LAE candidates.
We also show the distribution of Ly$\alpha$ luminosities for the full sample
(18 LAEs) of Rhoads \& Malhotra (2001).
Our LAE candidates appear to be more luminous than the
 LAE candidates found by Rhoads \& Malhotra (2001).
The limit of the Ly$\alpha$ luminosity for our survey,
 $L_{\rm lim}({\rm Ly}\alpha) \approx 5.0 \times 10^{42} h_{0.7}^{-2}$ergs s$^{-1}$,
 is larger than that of theirs,
 $L_{\rm lim}({\rm Ly}\alpha) \approx 2.5 \times 10^{42} h_{0.7}^{-2}$ergs s$^{-1}$.
This difference explains why our LAE candidates tend to be more luminous than
 those found by Rhoads \& Malhotra (2001).


\subsection{Star Formation Rates of the LAE Candidates}
\label{sec:sfr}
We estimate the SFRs for the 20 LAE candidates at $z \approx 5.7$.
Using equation (\ref{LHtoSFR}),
 together with a relation
\begin{equation}
\label{HtoLy}
L({\rm Ly}\alpha) = 8.7 L({\rm H}\alpha)
\end{equation}
from Case B recombination theory (Brocklehurst 1971),
we can estimate the SFRs using the Ly$\alpha$ luminosity;
\begin{equation}
SFR({\rm Ly}\alpha) = 9.1 \times 10^{-43} L({\rm Ly}\alpha) ~~~
M_\odot {\rm yr}^{-1}
\end{equation}
where, $L($Ly$\alpha)$ is in units
 of ergs s$^{-1}$ (see Hu et al. 1998).
The results are given in last column of Table \ref{tab:LLAE}.
We note that the SFR derived here is reduced
by the cosmic transmission.
The SFRs range from 5 to 14 $h_{0.7}^{-2} M_\odot$ yr$^{-1}$
with an average of $8.1 \pm 0.3 h_{0.7}^{-2} M_\odot$ yr$^{-1}$.
Although these values are typical to the Lyman break galaxies
at $z \sim$ 3 -- 4 (e.g., Steidel et al. 1999 and references therein),
the number density of the strong LAEs like our sources
is rather small.

\begin{figure}
\epsscale{0.47}
\plotone{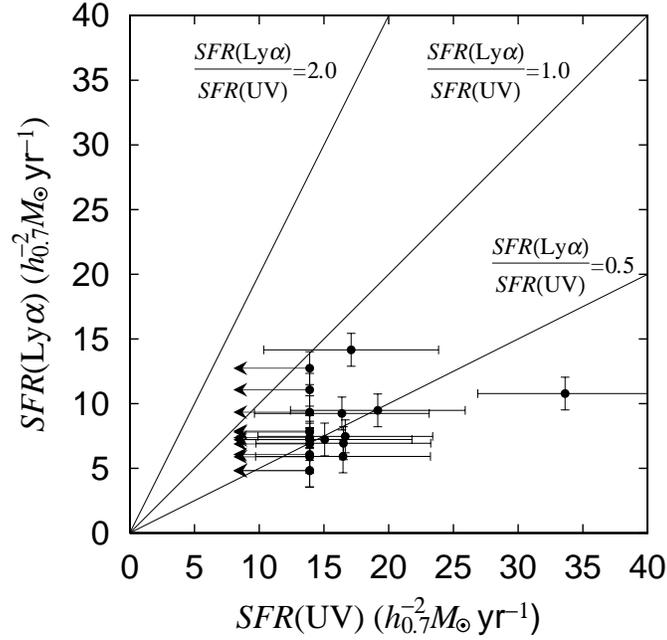}
\caption{Comparison between $SFR$(Ly$\alpha$) and $SFR$(UV)
         for the 20 LAE candidates.
\label{sfruv}}
\end{figure}

We examine whether or not the SFR derived from the Ly$\alpha$ luminosity
is consistent with that derived from the UV continuum luminosity
for our sample. The observed $z^\prime$ magnitude can be converted to
a UV continuum luminosity at $\lambda = 1340$ \AA.
Using the following relation (Kennicutt 1998; see also Madau et al. 1998),

\begin{equation}
\label{UVtoSFR}
SFR({\rm UV}) = 1.4 \times 10^{-28}L_{\nu}
~~ M_{\odot} ~ {\rm yr}^{-1},
\end{equation}
where $L_\nu$ is in units of ergs s$^{-1}$ Hz$^{-1}$,
we estimate the SFR based on the rest-frame UV ($\lambda=1340$\AA)
continuum luminosity for each object. The results are summarized
in Table \ref{tab:UV}.
Then we compare the two SFRs, $SFR$(Ly$\alpha$) and $SFR$(UV),
for each object in Figure \ref{sfruv}.
It is shown that $SFR$(UV) appears greater than $SFR$(Ly$\alpha$)
for most of our LAE candidates.
We also examine global ratio of two total SFRs,
 $SFR_{\rm total}$(Ly$\alpha$) / $SFR_{\rm total}$(UV)$ = 0.77^{+0.13}_{-0.10}$,
where $SFR_{\rm total}$(Ly$\alpha$) is the sum of $SFR$(Ly$\alpha$)
 of all our LAE candidates
and $SFR_{\rm total}$(UV) is that of $SFR$(UV).
If only 8 LAE candidates above the 2 $\sigma$ detection
in the $z^\prime$-band are taken into consideration, this ratio decreases to
$SFR_{\rm total}$(Ly$\alpha$) / $SFR_{\rm total}$(UV)$ = 0.47^{+0.07}_{-0.06}$.
The lower values of $SFR$(Ly$\alpha$) may be due to the effect of
 the absorption by the dust and scattering by the intergalactic
medium (see Hu et al. 2002).

\subsection{Star Formation Rate Density at $z\approx 5.7$}

It is interesting to estimate the contribution of the star formation
in the 20 LAE candidates to the co-moving
cosmic star formation rate density (SFRD; e.g., Madau et al. 1996).
Integrating the SFRs of the 20 LAE candidates, and correcting for
the detection completeness, we obtain the co-moving SFRD for our sources,
$\rho_{\rm SFR} \sim 1.2 \times 10^{-3} h_{0.7} M_\odot$ yr$^{-1}$ Mpc$^{-3}$.

\begin{figure}
\epsscale{0.47}
\plotone{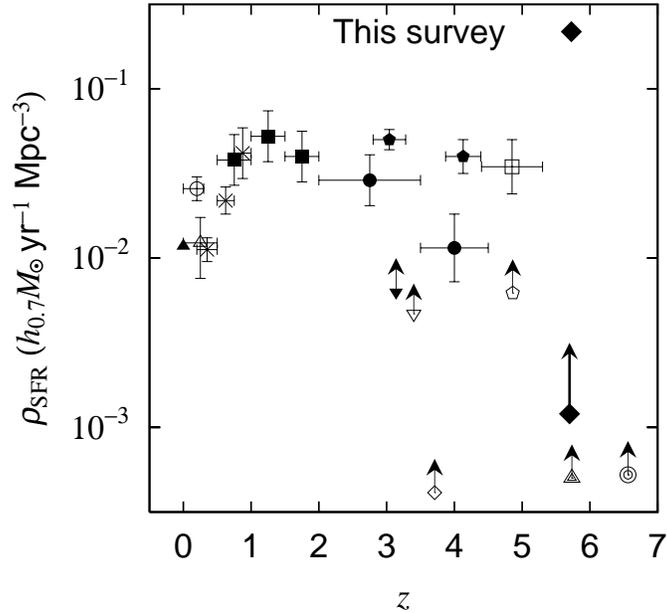}
\caption{Star formation rate density (SFRD) at $z \approx 5.7$ derived
         in this study (filled diamond)
         is shown together with the previous investigations.
         The other data sources are
         Gallego et al. (1996 -- filled triangle),
         Lilly et al. (1996 -- stars),
         Connolly et al. (1997 -- filled squares),
         Cowie \& Hu (1998 -- open inverse triangle),
         Madau et al. (1998 -- filled circles),
         Tresse \& Maddox (1998, open circle),
         Treyer et al. (1998 -- open triangle),
         Steidel et al. (1999 -- filled pentagons),
         Kudritzki et al. (2000 -- filled inverse triangle),
         Ouchi et al. (2003 -- open hexagons),
         Fujita et al. (2003a -- open diamond),
         Rhoads et al. (2003 -- double triangle),
         Kodaira et al. (2003 -- double circle), and
         Iwata et al. (2003 -- open square).
\label{sfrd}}
\end{figure}

In Figure \ref{sfrd}, we compare this SFRD with those
of previous studies compiled by Kodaira et al. (2003).
Note that the results of all surveys shown in this figure except for the LAE
 surveys are estimated by converting the H$\alpha$ or UV luminosity densities
 (obtained by integrating the LFs from $L=0$ to $\infty$) to the SFRDs by
 using the equation (\ref{LHtoSFR}) or (\ref{UVtoSFR}), respectively.
As shown in this figure, the SFRD derived in this study
is much smaller than the previous estimates based on the LBG surveys at $z= 3$ -- 5
(Madau et al. 1998; Steidel et al. 1999; and Iwata et al. 2003)
by an order of magnitude.

On the other hand, our result appears consistent with those of the previous LAE
surveys for redshift range between $z\sim3$ and $z\sim6$ 
 (Cowie \& Hu 1998; Kudritzki et al. 2000; Fujita et al. 2003a;
 Ouchi et al. 2003; Rhoads et al. 2003; Kodaira et al. 2003)
 within a factor of 5.
Note that all the values shown in Figure \ref{sfrd} are re-estimated
 with the adopted cosmology. Therefore, for readers' convenience,
 we give them in Table \ref{tab:pre}.

The SFRDs estimated from the LAE samples given in Table \ref{tab:pre}
 provide only lower limits because of the following reasons.
First, no correction is made for the extinction by dust and/or intergalactic gas.
To correct for these effects, we need much deeper optical and near-infrared
 observations (e.g. Hu et al. 2002).
Second, we do not integrate the SFRD
 of LAE candidates assuming a certain luminosity function from
 a lower to a upper limit because there is no
 reliable luminosity function for high-$z$ LAEs.
To estimate the reliable luminosity function for high-$z$ LAEs,
 we need much deeper wide-field surveys to obtain a large sample of LAEs.
Therefore, future careful investigations
 will be absolutely necessary to estimate a more reliable contribution
 of LAEs to the cosmic SFRD at high redshift.

For future consideration, we try to estimate a Ly$\alpha$ luminosity function (LF)
and then an integrated SFRD using the Ly$\alpha$ LF.
First, we assume that the Ly$\alpha$ LF has the same functional form as 
 that of H$\alpha$ LF because it is expected that the Ly$\alpha$ luminosity
 is correlated to the H$\alpha$ luminosity as given in equation (\ref{HtoLy}).
Here we use the H$\alpha$ LF at $z\approx0.24$ derived by
 Fujita et al. (2003b) which is described as a Schechter function
with $\alpha = -1.53 \pm 0.15$ and $\log\phi_* = -2.62 \pm 0.34$,
 where $\phi_*$ is in units of Mpc$^{-3}$;
note that their H$\alpha$ LF has $\log L_* = 41.95 \pm 0.25$,
 where $L_*$ is in units of ergs s$^{-1}$.
This parameter is treated as a free parameter in the following analysis.

\begin{figure}
\epsscale{0.47}
\plotone{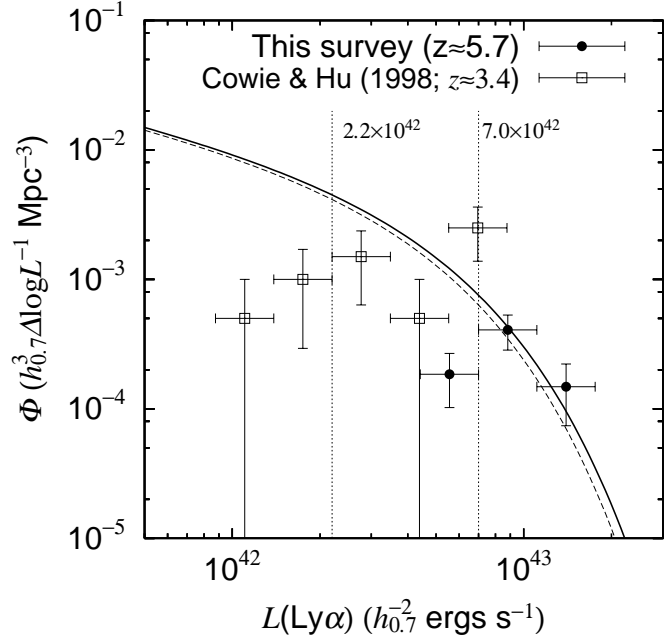}
\caption{Luminosity function (LF) of the our LAE candidates (solid curve) and
 that of Cowie \& Hu (1998 -- dashed curve).
 The vertical dotted lines show the complete limit, $L_{\rm lim}$, of our survey
 ($7.0 \times 10^{42} h_{0.7}^{-2}$ ergs s$^{-1}$) and that of Cowie \& Hu
 (1998; $2.2 \times 10^{42} h_{0.7}^{-2}$ ergs s$^{-1}$).
\label{lf}}
\end{figure}

Since the small luminosity range covered by our LAE candidates,
 we estimate $L_*$ for the Ly$\alpha$ LF satisfying the following equation
 instead of fitting the LF;
\begin{equation}
\mathcal{L}(L_{\rm lim})= \int^{\infty}_{L_{\rm lim}/L_*}
 L \phi_* \left(\frac{L}{L_*}\right)^\alpha
 \exp\left(-\frac{L}{L_*}\right)d\left(\frac{L}{L_*}\right),
\end{equation}
 where $\mathcal{L}(L_{\rm lim})$ is the Ly$\alpha$ luminosity density for
 the LAE candidates with $L({\rm Ly}\alpha)>L_{\rm lim}$ and $L_{\rm lim}$
 is the completeness limit of our survey
 ($L_{\rm lim} = 7.0 \times 10^{42} h_{0.7}^{-2}$ ergs s$^{-1}$
 corresponding to the Ly$\alpha$ luminosity of the LAE with ${\it NB816}=25.0$
 whose whole flux in the {\it NB816} passband is assumed to be attributed to
 the Ly$\alpha$-emission line).
Adopting the above Schechter function parameters ($\alpha=-1.53$ and
 $\log \phi_* = -2.62$), and using
 $\mathcal{L}(L_{\rm lim})=1.1 \pm 0.3 \times 10^{39} h_{0.7}$
 ergs s$^{-1}$ Mpc$^{-3}$
 for our 15 LAE candidates with $L({\rm Ly}\alpha)>L_{\rm lim}$,
 we find $L_*=4.1 \pm 0.4 \times 10^{42} h_{0.7}^{-2}$ ergs s$^{-1}$.
The Ly$\alpha$ LF is shown in Figure \ref{lf}.
Note that the detection completeness is taken into account in
the estimate of $\mathcal{L}(L_{\rm lim})$.

Using the same procedure described above, we also try to estimate
a Ly$\alpha$ LF at $z\approx3.4$ using the survey data of Cowie \& Hu (1998).
In this estimate, we use 
 $\mathcal{L}(L_{\rm lim})=4.6 \pm 1.7 \times 10^{39} h_{0.7}$ ergs s$^{-1}$ Mpc$^{-3}$
for the 9 objects with
 $L($Ly$\alpha) > L_{\rm lim} = 2.2 \times 10^{42} h_{0.7}^{-2}$ ergs s$^{-1}$.
It corresponds to the Ly$\alpha$ luminosity of the LAE with the limiting
 narrowband magnitude of their survey (HDF LRIS field),
 $m_{\rm AB}=25.0$, where we assumed that the whole flux in their narrowband
 ($\Delta\lambda=77$ \AA) imaging is attributed to the Ly$\alpha$-emission line.
We then find $L_*=3.8 \pm 0.8 \times 10^{42} h_{0.7}^{-2}$ ergs s$^{-1}$ and the
 Ly$\alpha$ LF is also shown in Figure \ref{lf} together
 the result for our LAE candidates at $z \approx 5.7$.
As shown in Figure \ref{lf}, there appears no evolution between
 $z\approx5.7$ and $z\approx3.4$, although the analyses given here are tentative.
Note that the detection completeness of the survey by Cowie \& Hu (1998)
 is not taken into account in the above estimate.

\begin{figure}
\epsscale{0.47}
\plotone{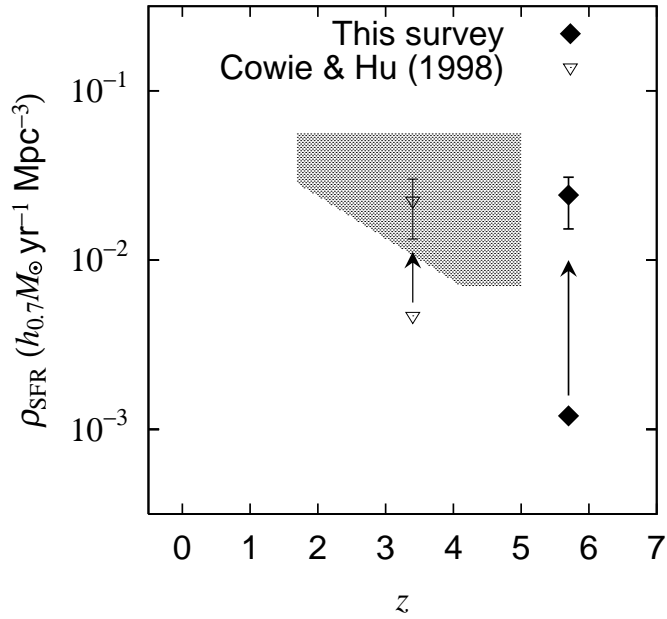}
\caption{Corrected SFRD at $z \approx 5.7$ derived
         in this study (filled diamond) is shown together 
         that of Cowie \& Hu (1998 -- open inverse triangle).
         The shaded area shows the distribution of SFRDs
         based on the LBG surveys (see Figure \ref{sfrd}).
\label{sfrdm}}
\end{figure}

Then, we re-estimate the SFRDs based on our survey and Cowie \& Hu (1998)
 integrating the Ly$\alpha$ LFs for the two surveys derived above.
In order to make it possible to compare these LF-corrected results with 
those of LBGs, we take account of the difference in the extinction
 between Ly$\alpha$ and UV continuum by adopting
 $SFR($Ly$\alpha)/SFR($UV$)=0.77^{+0.13}_{-0.10}$
(see section \ref{sec:sfr}; see also Hu et al. 2002).
As a result, we obtain a corrected SFRD at $z\approx5.7$, 
$\rho_{\rm SFR}^{\rm corr} \sim 2.4^{+0.6}_{-0.9}\times 10^{-2} h_{0.7} M_\odot$
 yr$^{-1}$ Mpc$^{-3}$,
where the uncertainty includes that in the estimate of the H$\alpha$ LF.
We also obtain a corrected SFRD for the LAEs  at $z\approx3.4$,
$\rho_{\rm SFR}^{\rm corr} \sim 2.3^{+0.8}_{-1.0}\times 10^{-2} h_{0.7} M_\odot$
 yr$^{-1}$ Mpc$^{-3}$.
These results are shown in Figure \ref{sfrdm}.
Although the lower limits of SFRDs at $z \approx 3.4$ and $z \approx 5.7$ 
appears significantly smaller than those based on the LBG surveys
(see Figure \ref{sfrd}), the LF-corrected SFRDs are comparable to those
based on the LBG surveys. If this is real,
the SFRD appears almost constant between $z \sim 1$ and $z \sim 6$.
Future detailed investigations are absolutely necessary to 
examine this result.

\subsection{Follow-up Optical Spectroscopy}
\label{follow}
Follow-up optical spectroscopy has been made of
two objects among the 20 LAE candidates;
(1) LAE J1044$-$0130 (Ajiki et al. 2002), and
(2) LAE J1044$-$0123 (Taniguchi et al. 2003); they are identified
as No. 18 and No. 17 in Table \ref{tab:LAE}, respectively.
These observations were made using ESI\footnote{
Echelle Spectrograph and Imager available on the Keck II telescope
(Sheinis et al. 2000).} on the Keck II Telescope and
FOCAS on the Subaru Telescope. The detailed information is
given in the above two papers.
These two objects are confirmed as an LAE at $z=5.687 \pm 0.002$
and $z=5.655\pm0.002$, respectively.
In Figure \ref{sp3}, we show their optical spectra obtained with ESI on the
 Keck II Telescope together with that of SDSSp J104433.04$-$012502.2
 obtained with FOCAS\footnote{Faint Object Camera And Spectrograph
 available on the Subaru Telescope (Kashikawa et al. 2002).}.

\begin{figure}
\epsscale{0.4}
\plotone{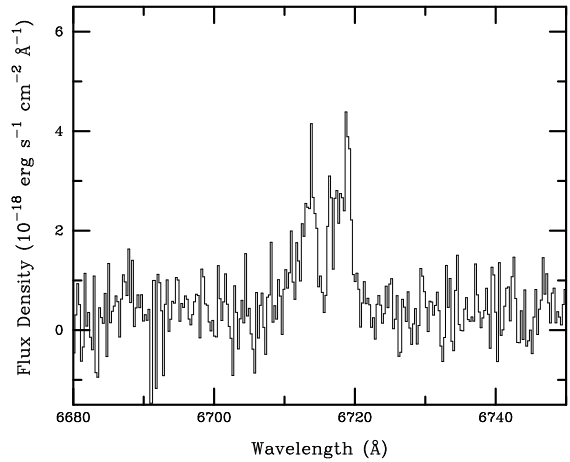}
\caption{
Optical spectrum of the foreground neighbor galaxy of
LAE J1044$-$0130 obtained with ESI on Keck II.
The \OII ~ doublet is clearly seen.
\label{oii}}
\end{figure}

During the course of our ESI spectroscopy of LAE J1044$-$0130,
we also obtained a spectrum of a galaxy seen 2.4 arcsec northeast
of LAE J1044$-$0130 (see No. 18 in Figure \ref{sum}).
We show this spectrum in Figure \ref{oii}.
Two emission lines are clearly seen at $\lambda \sim$
6710 -- 6720 \AA. These lines are identified as [O {\sc ii}]$\lambda$3727
doublet redshifted to $z=0.802 \pm 0.002$.
Most LAEs show a single emission line in their optical spectra
and thus it is often difficult to identify an LAE from other low-$z$
interlopers if the spectral resolution is not high enough to
check the emission-line profile in detail (e.g., Stern et al. 2000).
As demonstrated in Figure \ref{oii}, high-resolution spectroscopy (e.g.,
$R > 2000$ -- 3000) is very useful to omit low-$z$ [O {\sc ii}]
emitters.

\subsection{Optical Counterpart Search for the Lyman Limit System at
$z=5.72$}

Finally, we give comments on a possible counterpart of
 the Lyman Limit System (LLS) at $z=5.72$ reported by Fan et al. (2000).
As shown in Figure \ref{map}, the nearest LAE to the quasar is
LAE J1044$-$0123.
As described in detail in Taniguchi et al. (2003), this LAE
 is located at a projected distance of 4.45 Mpc from the quasar
 if its cosmological distance is the same as that of the quasar.
This LAE is too far from the quasar to be identified as
 a counterpart of the LLS.
Further the redshift of this LAE ($z=5.655$) is different from that of the LLS.
Although we find a faint galaxy with
 $m_B$(AB) $\approx 25$ is located at 1.9 arcsec southwest of the quasar,
 its broad band color properties from $B$ to $z^\prime$ suggest that
 the galaxy is a foreground galaxy
 located at a redshift of $z \sim 1.5$ -- 2.5 (Shioya et al. 2002).
In conclusion, we cannot find any possible counterpart of the LLS.

\section{SUMMARY}

We have presented our optical narrowband
({\it NB816}: $\lambda_{\rm C} = 8150$ \AA ~ and $\Delta\lambda = 120$ \AA)
observations of the field surrounding the high-$z$ quasar
 SDSSp J104433.04$-$012502.2 obtained with Suprime-Cam on the
8.2 m Subaru Telescope.
The survey area is 720 arcmin$^2$.
Our survey probes a volume of $1.8 \times 10^{5} h_{0.7}^{-3}$ Mpc$^{3}$.

(1) In our survey, we have found 62 emission-line sources whose
observed emission-line equivalent widths are greater than 180 \AA.
From their optical multicolor properties, we have identified 20 LAE
candidates at $z \approx$ 5.7 ($\Delta z \approx 0.1$).
If they were low-$z$ emitters, they would be very luminous star-forming
galaxies and thus would be easily detected in our $B$ and $R_{\rm C}$
images because their predicted apparent magnitudes are fairly brighter than
our limiting magnitudes.
Note also that two of them have been already confirmed from our follow-up
optical spectroscopy as LAEs
at $z=5.655$ (Taniguchi et al. 2003) and $z=5.687$ (Ajiki et al. 2002).

(2) We find that the rest-frame Ly$\alpha$ equivalent widths of the 20 LAE
candidates range from 25\AA ~ to $>120$\AA, being consistent with that of 
LAEs at $z>5$ (e.g., Dey et al. 1998; Hu et al. 2002; Dawson et al. 2002).

(3) After correcting for the detection
completeness, we obtained a space density of the LAE candidates,
$n({\rm Ly}\alpha) \simeq 1.5 \times 10^{-4} h_{0.7}^{3}$ Mpc$^{-3}$ at $z \approx 5.7$.
This density is higher by a factor of 2 than that of the
survey for LAEs at $z\approx5.7$ by Rhoads \& Malhotra (2001).

(4) We investigate the Ly$\alpha$ luminosities for the 20 LAE candidates.
The derived Ly$\alpha$ luminosities range from
$\approx 5 \times 10^{42}$ to $1.4 \times 10^{43} h_{0.7}^{-2}$ ergs s$^{-1}$.
Our survey probes higher-luminosity sources than that
 of Rhoads \& Malhotra (2001)
 since the equivalent width criterion of our survey
 ($EW_{\rm obs} \gtrsim 180$ \AA) is higher than that of Rhoads \&
Malhotra (2001; $EW_{\rm obs} > 80$ \AA).

(5) We obtain SFRs of our 20 LAE candidates ranging from 5 to 14
 $h_{0.7}^{-2} M_\odot$ yr$^{-1}$ with an average of
 $8.1 \pm 0.3 h_{0.7}^{-2} M_\odot$ yr$^{-1}$.
Although these values are typical to those of the Lyman break galaxies
at $z \sim$ 3 -- 4, the number density of the strong LAEs like our sources
appears rather small, since the present survey is not deep enough to detect
 faint emission line galaxies.
We also compare the two SFRs, $SFR$(Ly$\alpha$) and $SFR$(UV),
for the 20 LAE candidates.
It appears that $SFR$(UV) is higher than $SFR$(Ly$\alpha$)
for most of our LAE candidates.
This may be due to the effect of the absorption by the dust and scattering
 by the intergalactic medium for Ly$\alpha$ emission (see Hu et al. 2002).

(6) We have estimated the contribution of the star formation activity
 in the 20 LAE candidates to the co-moving
 cosmic SFRD (e.g., Madau et al. 1996).
The total star formation rate of our 20 LAE candidates with correction
 for the detection completeness gives
$\rho_{\rm SFR} \sim 1.2 \times 10^{-3} h_{0.7} M_\odot$ yr$^{-1}$ Mpc$^{-3}$.
This value
is much smaller than the previous estimates by LBG surveys at $z=3$ -- 5
(Madau et al. 1998; Steidel et al. 1999; Iwata et al. 2001) by an order of
 magnitude.
However, if we integrate the SFRD adopting a Ly$\alpha$ LF, the SFRD could be
 as high as those of the previous surveys for LBGs at $z\sim$ 3 -- 5.

\vspace{4ex}

We would like to thank both the Subaru and Keck Telescopes staff for
their invaluable help, and Dave Sanders, Sylvain Veilleux, S. Okamura,
Y. Ohyama, S. Oyabu, N. Kashikawa, M. Iye, H. Ando, and H. Karoji
 for encouragement during the course of this study.
We would also thank James Rhoads for providing us useful information on
 the LALA survey, M. Ouchi, M. Yagi, and K. Shimasaku for useful discussion
 on the data reduction of Suprime-Cam data,
 and T. Hayashino for his technical help.
We also thank an anonymous referee for his/her useful comments and suggestions.
This work was financially supported in part by
the Ministry of Education, Culture, Sports, Science, and Technology
(Nos. 10044052, and 10304013). 
MA and TN are JSPS fellows.



\begin{figure}
\begin{center}
\begin{tabular}{|c|}
\hline
ajiki\_fig2.jpg \\
\hline
\end{tabular}
\end{center}
\caption{Final {\it NB816} image (north is up).
         Our actual survey area is enclosed by solid line.
\label{nbg}}
\end{figure}
\begin{figure}
\begin{center}
\begin{tabular}{|c|}
\hline
ajiki\_fig3.png \\
\hline
\end{tabular}
\end{center}
\caption{Objects detected to the apparent magnitude limit of
         ${\it NB816}=26$ in the {\it NB816}-selected catalog.
         The objects selected as LAE candidates are shown as filled circles,
         and other emitters are shown as open circles.
         The horizontal solid line corresponds to the color of
         ${\it Iz-NB816}=0.9$. Objects above this line have strong
         emission lines with $EW_{\rm obs} \approx 180$ \AA ~ or greater.
         The dashed curves show the distribution of $3 \sigma$ error.
\label{nbcm}}
\end{figure}

\begin{figure}
\begin{center}
\begin{tabular}{|c|}
\hline
ajiki\_fig4.png \\
\hline
\end{tabular}
\end{center}
\caption{Diagram of objects detected above 3$\sigma$ in both $R_{\rm C}$-,
         $I_{\rm C}$- and $z^\prime$-band image
         in the {\it NB816} selected catalog
         between $R_{\rm C}-I_{\rm C}$ and $I_{\rm C}-z^\prime$ (right
         panel).
         In the left panel, we show the color-color diagrams expected for
         five typical type of galaxies(see text).
         The horizontal line shows selection criteria of LAE
         candidates with $I_{\rm C}\leq24.8$.
\label{riz}}
\end{figure}
\begin{figure}
\begin{center}
\begin{tabular}{|c|}
\hline
ajiki\_fig5a.png \\
\hline
ajiki\_fig5b.png \\
\hline
ajiki\_fig5c.png \\
\hline
ajiki\_fig5d.png \\
\hline
\end{tabular}
\end{center}
\caption{Broad-band and {\it NB816} images (upper panels) and
          contour maps (middle panels) of our
          LAE candidates at $z \approx 5.7$.
         Each box is $16^{\prime \prime}$ on a side.
         Each circle is $4^{\prime \prime}$ radius.
         The contours show surface brightness of 25, 26 and 27 mag
arcsec$^{-2}$.
         The SED is also shown in lower panel for each object.
\label{sum}}
\end{figure}







\clearpage

\begin{deluxetable}{lcccc}
\tablenum{1}
\tablecaption{Journal of observations\label{tab:obs}}
\tablewidth{0pt}
\tablehead{
\colhead{Band} &
\colhead{Obs. Date (UT)} &
\colhead{$T_{\rm int}$ (sec)\tablenotemark{a}}  &
\colhead{$m_{\rm lim}$(AB)\tablenotemark{b}} &
\colhead{$FWHM_{\rm star}$ (arcsec)\tablenotemark{c}}
}
\startdata
$B$            & 2002 February 17      &  1680 & 26.6 & 1.2 \\
$R_{\rm C}$    & 2002 February 15, 16  &  4800 & 26.2 & 1.4 \\
$I_{\rm C}$    & 2002 February 15, 16  &  3360 & 25.9 & 1.2 \\
$z'$           & 2002 February 15, 16  &  5160 & 25.3 & 1.2 \\
${\it NB816}$  & 2002 February 15 - 17 & 36000 & 26.0 & 0.9 \\
\enddata
\tablenotetext{a}{Total integration time.}
\tablenotetext{b}{The limiting magnitude (3$\sigma$) within a
2.8 arcsec aperture.}
\tablenotetext{c}{The full width at half maximum of stellar
objects in the final image}
\end{deluxetable}

\begin{deluxetable}{rccrrrrrr}
\tablenum{2}
\tablecaption{Photometric properties of the LAE candidates at $z
\approx 5.7$ \tablenotemark{a} \label{tab:LAE}}
\tablewidth{0pt}
\tablehead{
\colhead{No.} &
\colhead{$\alpha$(J2000)} &
\colhead{$\delta$(J2000)} &
\colhead{$B$\tablenotemark{a}} &
\colhead{$R_{\rm C}$\tablenotemark{a}} &
\colhead{$I_{\rm C}$\tablenotemark{a}} &
\colhead{${\it NB816}$\tablenotemark{a}} &
\colhead{$z'$\tablenotemark{a}} &
\colhead{${\it Iz}$\tablenotemark{a}} \\
\colhead{} &
\colhead{h ~~ m ~~ s} &
\colhead{$^\circ$ ~~ $^\prime$ ~~ $^{\prime\prime}$}  &
\colhead{}&
\colhead{}&
\colhead{}&
\colhead{}&
\colhead{}&
\colhead{}
}
\startdata
 1 & 10 43 31.9 & $-$01 14 46 & $>$27.0 & $>$26.6 &  (26.2) & 24.6 &  (25.5) &    26.0 \\
 2 & 10 43 38.6 & $-$01 14 06 & $>$27.0 & $>$26.6 & $>$26.3 & 24.5 &  (25.5) &  (26.3) \\
 3 & 10 43 43.0 & $-$01 27 43 & $>$27.0 & $>$26.6 & $>$26.3 & 24.7 & $>$25.7 & $>$26.5 \\
 4 & 10 43 44.4 & $-$01 15 59 & $>$27.0 & $>$26.6 &    25.5 & 24.1 &    24.8 &    25.3 \\
 5 & 10 43 44.5 & $-$01 24 31 & $>$27.0 & $>$26.6 &  (25.9) & 24.7 &  (25.5) &    25.8 \\
 6 & 10 43 45.6 & $-$01 43 20 & $>$27.0 & $>$26.6 &  (26.2) & 24.3 &  (25.4) &    25.9 \\
 7 & 10 43 48.3 & $-$01 23 20 & $>$27.0 & $>$26.6 &    25.7 & 23.9 &  (25.5) &    25.6 \\
 8 & 10 43 53.0 & $-$01 36 12 & $>$27.0 & $>$26.6 &  (26.2) & 25.0 & $>$25.7 &  (26.2) \\
 9 & 10 43 55.0 & $-$01 21 48 & $>$27.0 & $>$26.6 & $>$26.3 & 24.7 &  (25.5) &  (26.1) \\
10 & 10 43 55.3 & $-$01 28 47 & $>$27.0 & $>$26.6 &  (26.1) & 24.5 & $>$25.7 &    26.0 \\
11 & 10 43 55.4 & $-$01 13 39 & $>$27.0 & $>$26.6 & $>$26.3 & 24.7 &  (25.6) & $>$26.5 \\
12 & 10 43 55.5 & $-$01 14 18 & $>$27.0 & $>$26.6 &  (26.2) & 25.0 & $>$25.7 &  (26.2) \\
13 & 10 43 59.0 & $-$01 16 27 & $>$27.0 & $>$26.6 &  (26.0) & 24.3 & $>$25.7 &  (26.2) \\
14 & 10 44 08.5 & $-$01 24 35 & $>$27.0 & $>$26.6 & $>$26.3 & 24.2 & $>$25.7 & $>$26.0 \\
15 & 10 44 10.2 & $-$01 16 18 & $>$27.0 & $>$26.6 & $>$26.3 & 24.9 & $>$25.7 & $>$26.0 \\
16 & 10 44 11.3 & $-$01 25 12 & $>$27.0 & $>$26.6 &  (26.1) & 24.7 & $>$25.7 & $>$26.0 \\
17\tablenotemark{b}
   & 10 44 26.6 & $-$01 23 45 & (26.8)  & $>$26.6 &  (25.9) & 24.7 & $>$25.7 &    25.9 \\
18\tablenotemark{c}
   & 10 44 32.4 & $-$01 30 34 & $>$27.0 & $>$26.6 &  (26.1) & 24.4 & $>$25.7 &    26.0 \\
19 & 10 44 41.4 & $-$01 38 30 & $>$27.0 & $>$26.6 & $>$26.3 & 25.0 & $>$25.7 & $>$26.0 \\
20 & 10 44 52.9 & $-$01 39 14 & $>$27.0 & $>$26.6 &  (26.1) & 24.8 & $>$25.7 & $>$26.0 \\
\enddata
\tablenotetext{a}{AB magnitude in a 2.8 arcsec diameter.
                  The magnitudes between the 2 $\sigma$ and 3 $\sigma$ detection
                  levels are put in parentheses.}
\tablenotetext{b}{LAE J1044$-$0123 (Taniguchi et al. 2003).}
\tablenotetext{c}{LAE J1044$-$0130 (Ajiki et al. 2002).}
\end{deluxetable}

\begin{deluxetable}{cccccc}
\tablenum{3}
\tablecaption{The SFRs assuming the LAE candidates are low-$z$ emitters
              \label{tab:CONT}}
\tablewidth{0pt}
\tablehead{
\colhead{Line} &
\colhead{$z_{\rm c}$\tablenotemark{a}} &
\colhead{$V$\tablenotemark{b}} &
\colhead{$E(B-V)$} &
\colhead{($L_{\rm min}$, $L_{\rm max}$)\tablenotemark{c}} &
\colhead{($SFR_{\rm min}$, $SFR_{\rm max}$)\tablenotemark{d}} \\
\colhead{} &
\colhead{} &
\colhead{($10^{4} h_{0.7}^{-3}$ Mpc$^{3}$)} &
\colhead{} &
\colhead{($10^{41} h_{0.7}^{-2}$ ergs s$^{-1}$)} &
\colhead{($h_{0.7}^{-2} M_\odot$ yr$^{-1}$)}}
\startdata
          &      &      &  0  & (0.018, 0.054) & (0.015, 0.043) \\
H$\alpha$ & 0.24 & 0.39 & 0.3 & (0.030, 0.089) & (0.024, 0.70) \\
          &      &      & 0.6 & (0.049, 0.146) & (0.039, 0.12) \\
\tableline
          &      &      & 0   &  (0.25, 0.73)  &  (0.25, 0.73) \\
\OIII     & 0.63 & 2.62 & 0.3 &  (0.60, 1.8)   &  (0.60, 1.8) \\
          &      &      & 0.6 &   (1.4, 4.2)   &   (1.4, 4.2) \\
\tableline
          &      &      &  0  &   (1.2, 3.5)   &   (1.1, 3.1) \\
\OII      & 1.19 & 7.30 & 0.3 &   (4.5, 13)    &   (4.0, 12) \\
          &      &      & 0.6 &    (17, 50)    &    (15, 44) \\
\enddata
\tablenotetext{a}{The central redshift corresponding to the central
                  wavelength of the {\it NB816} filter.}
\tablenotetext{b}{The commoving volume covered by the {\it NB816} filter.}
\tablenotetext{c}{The minimum and maximum line luminosity, if our LAE
                  candidates are the corresponding line emitters.}
\tablenotetext{d}{The minimum and maximum SFR, if our LAE candidates are the
                  corresponding line emitters.}
\end{deluxetable}

\begin{deluxetable}{rccc}
\tablenum{4}
\tablecaption{Ly$\alpha$ luminosity and star formation rate for the
LAE candidates at $z \approx 5.7$\label{tab:LLAE}}
\tablewidth{0pt}
\tablehead{
\colhead{No.} &
\colhead{$EW_0$} &
\colhead{$L({\rm Ly}\alpha)$\tablenotemark{a}} &
\colhead{$SFR$(Ly$\alpha$)\tablenotemark{b}}  \\
\colhead{} &
\colhead{(\AA)} &
\colhead{($10^{43} h_{0.7}^{-2}$ ergs s$^{-1}$)} &
\colhead{($h_{0.7}^{-2} M_\odot$ yr$^{-1}$)}
}
\startdata
 1 &   63$^{+62}_{-25}$  & 0.8 &  7 \\
 2 & $>$80               & 1.0 &  9 \\
 3 & $>$52               & 0.8 &  7 \\
 4 &   46$^{+18}_{-12}$  & 1.2 & 11 \\
 5 &   40$^{+31}_{-16}$  & 0.7 &  6 \\
 6 &   81$^{+75}_{-31}$  & 1.0 &  9 \\
 7 &  100$^{+66}_{-32}$  & 1.6 & 14 \\
 8 & $>$36               & 0.5 &  5 \\
 9 & $>$57               & 0.8 &  7 \\
10 & $>$72               & 0.9 &  8 \\
11 & $>$51               & 0.8 &  7 \\
12 & $>$37               & 0.5 &  5 \\
13 & $>$110              & 1.2 & 11 \\
14 & $>$118              & 1.4 & 13 \\
15 & $>$43               & 0.9 &  8 \\
16 & $>$58               & 0.8 &  7 \\
17 &   47$^{+90}_{-19}$  & 0.6 &  6 \\
18 &   90$^{+137}_{-36}$ & 1.0 &  9 \\
19 & $>$35               & 0.8 &  7 \\
20 & $>$44               & 0.7 &  6 \\
\enddata
\tablenotetext{a}{$\sigma(L) \approx 0.1 \times 10^{43} h_{0.7}^{-2}$ ergs
s$^{-1}$}
\tablenotetext{b}{$\sigma(SFR) \approx 1~h_{0.7}^{-2} M_\odot$ yr$^{-1}$}
\end{deluxetable}

\clearpage
\begin{deluxetable}{rcccc}
\tablenum{5}
\tablecaption{UV continuum luminosity and star formation rate for the
LAE candidates at $z \approx 5.7$\label{tab:UV}}
\tablewidth{0pt}
\tablehead{
\colhead{No.} &
\colhead{$f_\nu$($z^\prime$)\tablenotemark{a}} &
\colhead{$L_{1340}$\tablenotemark{b}} &
\colhead{$SFR$(UV)\tablenotemark{c}} &
\colhead{$SFR({\rm Ly\alpha})/SFR({\rm UV})$} \\
\colhead{} &
\colhead{($10^{-30}$ ergs s$^{-1}$ cm$^{-2}$ Hz$^{-1}$)} &
\colhead{($10^{29} h_{0.7}^{-2} $ ergs s$^{-1}$ Hz$^{-1}$)} &
\colhead{($h_{0.7}^{-2} M_\odot$ yr$^{-1}$)} &
\colhead{}
}
\startdata
 1 &    2.3 &    1.2  &    17 &    0.44 \\
 2 &    2.2 &    1.2  &    16 &    0.55 \\
 3 & $<$1.9 & $<$1.0  & $<$14 & $>$0.52 \\
 4 &    4.5 &    2.4  &    34 &    0.32 \\
 5 &    2.2 &    1.2  &    16 &    0.36 \\
 6 &    2.6 &    1.4  &    19 &    0.50 \\
 7 &    2.3 &    1.2  &    17 &    0.82 \\
 8 & $<$1.9 & $<$1.0  & $<$14 & $>$0.34 \\
 9 &    2.2 &    1.2  &    17 &    0.42 \\
10 & $<$1.9 & $<$1.0  & $<$14 & $>$0.56 \\
11 &    2.0 &    1.1  &    15 &    0.48 \\
12 & $<$1.9 & $<$1.0  & $<$14 & $>$0.34 \\
13 & $<$1.9 & $<$1.0  & $<$14 & $>$0.79 \\
14 & $<$1.9 & $<$1.0  & $<$14 & $>$0.91 \\
15 & $<$1.9 & $<$1.0  & $<$14 & $>$0.56 \\
16 & $<$1.9 & $<$1.0  & $<$14 & $>$0.52 \\
17 & $<$1.9 & $<$1.0  & $<$14 & $>$0.42 \\
18 & $<$1.9 & $<$1.0  & $<$14 & $>$0.67 \\
19 & $<$1.9 & $<$1.0  & $<$14 & $>$0.49 \\
20 & $<$1.9 & $<$1.0  & $<$14 & $>$0.43 \\
\enddata
\tablenotetext{a}{$\sigma(f_\nu) \approx 0.9 \times 10^{-30}$ ergs s$^{-1}$
                  cm$^{-2}$ Hz$^{-1}$.}
\tablenotetext{b}{The UV continuum luminosity at $\lambda=1340$ \AA.}
\tablenotetext{c}{$\sigma(SFR) \approx 7~h_{0.7}^{-2} M_\odot$ yr$^{-1}$}
\end{deluxetable}

\clearpage
\begin{deluxetable}{rccc}
\tablenum{6}
\tablecaption{The SFRDs derived from previous studies\label{tab:pre}}
\tablewidth{0pt}
\tablehead{
\colhead{No.} &
\colhead{$(z_{\rm min}, z_{\rm max})$\tablenotemark{a}} &
\colhead{$\rho_{\rm SFR}$} &
\colhead{Reference\tablenotemark{b}}\\
\colhead{} &
\colhead{} &
\colhead{($h_{0.7} M_\odot$ yr$^{-1}$ Mpc$^{-3}$)}&
\colhead{}
}
\startdata
\multicolumn{4}{c}{The H$\alpha$-emitter surveys\tablenotemark{c}} \\
\tableline
1 & (0.000, 0.045) & $1.2\times 10^{-2}$ & 1 \\
2 & (0.0, 0.3)   & $2.5\times 10^{-2}$ & 2 \\
\tableline
\multicolumn{4}{c}{The multi color surveys\tablenotemark{d}} \\
\tableline
 1 & (0.0, 0.5)  & $1.2\times 10^{-2}$ & 3 \\
 2 & (0.20, 0.50) & $1.1\times 10^{-2}$ & 4 \\
 3 & (0.50, 0.75) & $2.2\times 10^{-2}$ & 4 \\
 4 & (0.5, 1.0)  & $4.2\times 10^{-2}$ & 5 \\
 5 & (0.75, 1.00) & $4.2\times 10^{-2}$ & 4 \\
 6 & (1.0, 1.5)  & $5.8\times 10^{-2}$ & 5 \\
 7 & (1.5, 2.0)  & $4.4\times 10^{-2}$ & 5 \\
 8 & (2.0, 3.5)  & $2.9\times 10^{-2}$ & 6 \\
 9 & (2.80, 3.28) & $5.0\times 10^{-2}$ & 7 \\
10 & (3.5, 4.5)  & $1.1\times 10^{-2}$ & 6 \\
11 & (3.87, 4.39) & $4.0\times 10^{-2}$ & 7 \\
12 & (4.4, 5.3) & $3.5\times 10^{-2}$ & 8 \\
\tableline
\multicolumn{4}{c}{The LAE surveys} \\
\tableline
 1 & (3.12, 3.15) & $6.4\times 10^{-3}$ & 9 \\
 2 & (3.41, 3.47) & $4.7\times 10^{-3}$ & 10 \\
 3 & (3.60, 3.83) & $4.1\times 10^{-4}$ & 11 \\
 4 & (4.83, 4.89) & $6.3\times 10^{-3}$\tablenotemark{e} & 12 \\
 5 & (5.65, 5.75) & $1.2\times 10^{-3}$ & This paper\\
 6 & (5.68, 5.80) & $5\times 10^{-4}$ & 13 \\
 7 & (6.51, 6.62) & $5.2\times 10^{-4}$ & 14 \\
\enddata
\tablenotetext{a}{The redshift range of each survey.}
\tablenotetext{b}{
1. Gallego et al. (1996),
2. Tresse \& Madox (1998),
3. Treyer et al. (1998),
4. Lilly et al. (1996),
5. Connolly et al. (1997),
6. Madau et al. (1998),
7. Steidel et al. (1999),
8. Iwata et al. (2003),
9. Kudritzki et al. (2000),
10. Cowie \& Hu (1998),
11. Fujita et al. (2003a); note that
    $\rho_{\rm SFR} \simeq 1 \times 10^{-3} h_{0.7} M_\odot$ yr$^{-1}$
    Mpc$^{-3}$ when we include their marginal and unclassified samples, 
12. Ouchi et al. (2003),
13. Rhoads et al. (2003) for their full sample (18 LAEs); note that 
    $\rho_{\rm SFR} \simeq 3 \times 10^{-4} h_{0.7} M_\odot$ yr$^{-1}$
    Mpc$^{-3}$ for their reduced sample (13 LAEs), \&
14. Kodaira et al. (2003).
}
\tablenotetext{c}{These SFRs are estimated by integrating the extinction-corrected
                  H$\alpha$ LFs from $L=0$ to $\infty$ with use of equation (\ref{LHtoSFR}).}
\tablenotetext{d}{These SFRs are estimated by integrating the
                  UV LFs from $L=0$ to $\infty$ with use of equation (\ref{UVtoSFR});
                  no correction for dust extinction is made.}
\tablenotetext{e}{Estimated from the UV continuum of LAE candidates.}
\end{deluxetable}
\end{document}